\documentclass[pra,twocolumn,showpacs]{revtex4-1}

\usepackage{graphicx,epstopdf}
\usepackage{amsmath}
\usepackage{amssymb}
\usepackage{color}
\usepackage{float}
\usepackage{epsfig}
\usepackage{epstopdf}
\usepackage{subfigure}

\usepackage[colorlinks=true, citecolor=blue, urlcolor=blue ]{hyperref}

\begin{document}

\title{Detecting phase boundaries of quantum spin-1/2 \emph{XXZ} ladder\\
via bipartite and multipartite entanglement transitions}

\author{Sudipto Singha Roy,\(^{1,2}\) Himadri Shekhar Dhar,\(^{1,2,3}\) Debraj Rakshit,\(^{1,2,4}\) Aditi Sen(De),\(^{1,2}\) and Ujjwal Sen\(^{1,2}\)}

\affiliation{\(^1\)Harish-Chandra Research Institute, Chhatnag Road, Jhunsi, Allahabad 211 019, India \\
\(^2\)Homi Bhabha National Institute, Anushaktinagar, Mumbai, Maharashtra 400 094, India \\
\(^3\)Institute for Theoretical Physics, Vienna University of Technology, Wiedner Hauptstra{\ss}e 8-10/136, A-1040 Vienna, Austria\\
\(^4\)Institute of Physics, Polish Academy of Sciences, Aleja Lotnik{\'o}w 32/46, PL-02668 Warsaw, Poland}

\date{\today}

\begin{abstract}
Phase transition in quantum many-body systems inevitably causes changes in certain physical properties which then serve as potential indicators of critical phenomena. Besides the traditional order parameters, characterization of quantum entanglement has  proven to be a computationally efficient and successful method for detection of phase boundaries, especially in one-dimensional models.
Here we determine  the rich phase diagram of the ground states of a quantum spin-1/2 \emph{XXZ}  ladder by analyzing 
the variation of bipartite and multipartite entanglements.
Our study characterizes the different ground state phases and notes the correspondence with known results, while highlighting the finer details that emerge from the behavior of ground state entanglement. Analysis of entanglement in the ground state provides a clearer picture of the complex ground state phase diagram of the system using only a moderate-size model.
\end{abstract}

\maketitle

\section{\label{intro} Introduction}

Quantum phase transition \cite{sachdev}, triggered by quantum fluctuations, is a rich cooperative phenomenon exhibited by many-body quantum systems at absolute zero temperature.
When a physical system undergoes a phase transition, some of its physical properties such as ground state energy, magnetization, classical correlation length, and quantum correlations, also change their pattern
and hence can serve as potential
indicators of the phase transition boundary \cite{sondhi, golden, sachdev,mojta}.
Incidentally, the challenging task in understanding the behavior of important many-body and strongly-correlated quantum systems is the characterization of the ground state phase boundaries using a computable order parameter. In such an enterprise, there are two main difficulties: firstly, there are only a handful of models that can be solved analytically and hence, studying the energy spectrum and eigenstates are only possible, in most cases, using approximate or numerical methods, which are often limited by size of the system. Secondly, there may not  exist a universal detector that can identify all the phase boundaries in a strongly-correlated system. To overcome these difficulties a possible way is to study these non-integrable models through different physical quantities and establish whether the outcomes using different techniques allow a phase structure to emerge.

In recent years, attempts have been made to explore quantum critical phenomena using properties which are beyond the traditional order parameters and, in some cases, are more efficient to compute, providing possibly deeper insight about the nature of correlations present in the system. Quantum correlation,
as characterized by entanglement \cite{horodecki}, is one such quantity.
Though entanglement
has initially been shown to be
a resource 
for quantum communication and other information-theoretic protocols \cite{nielsen, wilde},
%
it also turns out to be an important tool in investigating cooperative phenomena in many-body models \cite{fazio, sir-mam}.
The fact that the ground state of an interacting many-body system is potentially entangled, 
and the distribution of its entanglement is expected to demonstrate distinct characteristics across possible phase boundaries, shedding new light on quantum criticality,
is a primary motivation to characterize and analyze the entanglement present in many-body systems \cite{fazio, sir-mam}.
Consequently, sincere efforts to build a conceptual bridge between quantum information and many-body physics have gradually developed in recent years.
This is primarily motivated by the fact that such physical systems are potential substrates to implement quantum information protocols in a realizable setting \cite{greiner}.
In Refs.~\cite{fazio-nature, osborne}, it was shown that the derivative of the nearest-neighbor bipartite entanglement, as measured using concurrence \cite{wooters}, can remarkably identify the quantum phase transition point in the quantum transverse Ising and \emph{XY} spin-1/2 chains.
Down the avenue, several works have also reported the successful detection of phase transition points using various bipartite \cite{PT_entanglement} and multipartite \cite{multi-ent} entanglement measures (cf.~\cite{fazio, multi-bi}), including topological order using entanglement entropy \cite{topo-ent, wen}. In recent years, information-theoretic measures of quantum correlation \cite{modi} have also been used to investigate cooperative phenomena in many-body systems \cite{discord-qpt}. However, most of the studies using these measures have been restricted to one-dimensional models.

In this work, we go beyond some of these limitations, and consider
an important class of physical systems, the interacting quantum spin-1/2 ladders, with extremely rich ground state properties. These models have gained a lot of attention, in recent years, through the several proposals that connect these systems to gapped spin Hamiltonians \cite{spin_gap} and exotic phenomena such as high $T_c$ superconductivity \cite{high_tc}.
Also, multi-leg spin-1/2 ladders have been extensively used to investigate Haldane's conjecture \cite{haldane} with respect to the dichotomy between quantum chains with integer and half-odd integer spins, and their gapped and gapless energy spectra, respectively \cite{conjec-work}.
However, though significant efforts have been made towards understanding the complex phase
diagram of spin ladder systems, successful characterization of the different quantum phases have only been achieved in some limiting cases or through approximate models \cite{conjec-work, strong, watanbe, hijii,zanardi}.

In this work, we investigate the phase diagram of the quantum spin-1/2 \textit{XXZ} ladder by analyzing the transition boundaries arising from the variation of bipartite and multipartite entanglement in the ground states of the system with moderately large number of spins. For these systems,
conventional order parameters such as spin correlation functions or spectral energy gap does not provide distinct
phase boundaries, and are often limited to highlighting the characteristic features of the phases at the limiting values.
In contrast, we observe that the behavior of bipartite entanglement between nearest neighbor (NN) spins along the rungs and legs and the variation of the genuine multiparty entanglement in the ground state of the quantum spin-1/2 ladder is able to identify even the less prominent phase boundaries of the system. Our work shows that entanglement can be a good figure of merit in investigating critical phenomena even in complex ground state phases of strongly-correlated systems.\

The paper is arranged as follows. In Sec.~\ref{model}, we begin with a discussion on the spin-1/2 \emph{XXZ} ladder Hamiltonian and its ground state phases.
In Sec.~\ref{properties}, we define the different physical quantities associated with the detection of phase boundaries, and their variation with the system parameters of the ladder Hamiltonian. In Sec.~\ref{Phase}, a qualitative study of the different emergent phases have been carried out and a schematic phase diagram of the system
is proposed. In Sec.~\ref{ferro}, we study the different phases of the spin-1/2 \emph{XXZ} ladder with ferromagnetic couplings along the legs.
The scaling of the energy gap with system size is compared with that of the genuine multipartite entanglement in Sec.~\ref{scale}. Finally, we conclude in Sec.~{\ref{conclusion}}.

\begin{figure}[t]
\epsfig{figure =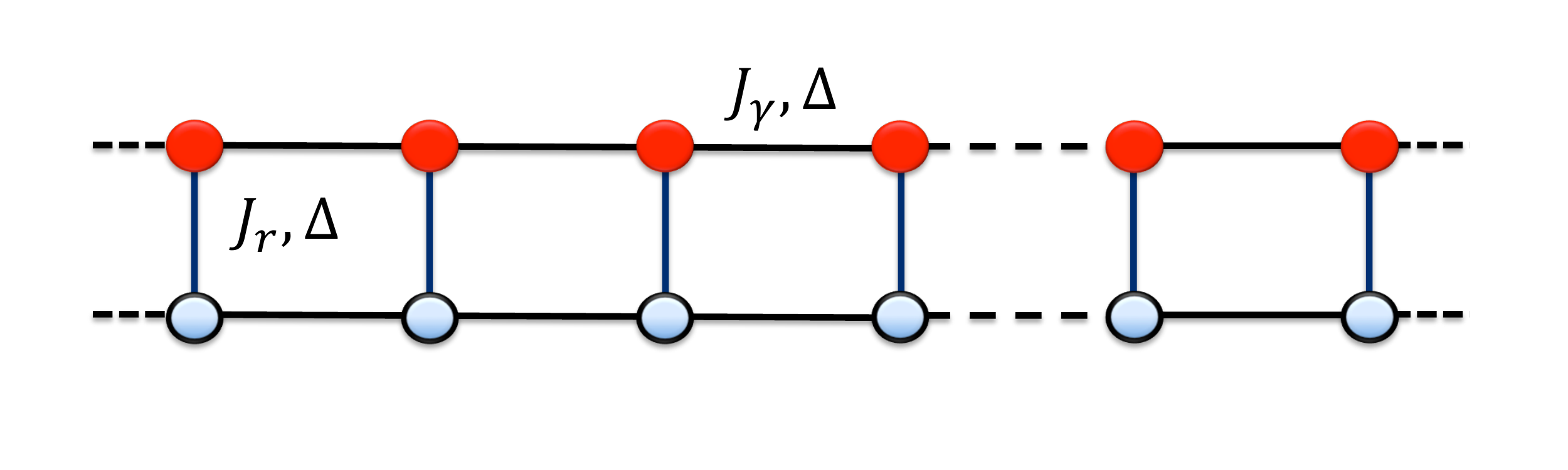, width=0.45\textwidth,angle=-0}
\caption{(Color online) A schematic representation of the \emph{XXZ} ladder. The figure shows a schematic diagram of a spin-1/2 \emph{XXZ} ladder, with couplings $J_\gamma$ and $J_r$ along the legs and rungs, respectively. The coupling constant for the interaction in the $z$-direction is $J_{\gamma}\Delta$ and $J_r \Delta$, respectively along the legs, and the rungs. The dotted lines at the end show the boundary condition  of the ladder.}
\label{con_leg}
\end{figure}
%

\section{\label{model} The Model}
Most studies on quantum spin ladders \cite{conjec-work, strong, watanbe, hijii,zanardi} consider antiferromagnetic coupling along the legs 
due to an interest in understanding exotic quantum phenomena, such as the experimentally observed high-$T_c$ transitions in antiferromagnetic ladder models for strontium-based superconducting cuprates \cite{high_tc, karen}.
Quantum spin-1/2 ladders with ferromagnetic interaction along the legs have  also received attention \cite{f1,f2,f3,f4}, particularly in the study of synthetic ferromagnetic chains built from Cu$^{2+}$ compounds \cite{f1}.
In our work, the primary focus is on quantum spin-1/2 \emph{XXZ} ladders with antiferromagnetic legs, given by the Hamiltonian $H=H_{l}+H_{r}$, where
\begin{eqnarray}\label{hamiltonian}
H_{l}&=&\sum_{\gamma=1}^2 J_{\gamma } \big(\sum_{i=1}^{N/2} S^x_{i,\gamma}S^x_{i+1,\gamma} + S^y_{i,\gamma}S^y_{i+1,\gamma}+
\Delta~ S^z_{i,\gamma}S^z_{i+1,\gamma} \big),\nonumber\\
H_{r}&=& J_{r} \big(\sum_{i'=1}^{N/2} S^x_{i',1}S^x_{i',2} + S^y_{i',1}S^y_{i',2} + \Delta~ S^z_{i',1}S^z_{i',2} \big), 
\end{eqnarray}
correspond to the interaction along the legs and rungs, respectively.
Here, $S^j$'s are the Pauli spin operators ($j$ = ${x,y,z}$), $\Delta$ is referred to as an anisotropy constant, differentiating the strength of interaction in the $z$-direction with those in the other two directions for legs as well as rungs, $J_\gamma>$ 0 is the antiferromagnetic coupling along the legs, and $J_{r}$ is the inter-leg or rung coupling.
We note that the coupling along the rungs can be either ferromagnetic or antiferromagnetic. Moreover, we consider periodic boundary conditions.
For completeness, in Sec.~\ref{ferro}, we also discuss the quantum phases of \emph{XXZ} ladders with ferromagnetic legs ($J_{\gamma}<$ 0) to show that
entanglement properties can again successfully detect
critical phenomena in these systems.

The extremely rich quantum phases of quantum spin-1/2 \emph{XXZ} ladders, with antiferromagnetic legs, are obtained by varying the $J_r/J_\gamma$ ratio and the anisotropy $\Delta$,
using numerical methods such as Abelian bosonization \cite{strong}, renormalization group \cite{watanbe}, and level spectroscopy \cite{hijii, nomura}, along with exact solutions at asymptotic parameter limits.
Recent numerical investigations have used sublattice entropy of entanglement in the XXZ ladder to detect specific phase boundaries \cite{zanardi}.
From these studies,
some agreement on the prominent phases have been achieved, although the exact demarcation of the different phases remains elusive \cite{hijii}.
For limiting values of the parameters $J_r/J_\gamma$  and $\Delta$, a coarse picture of some of the phases of the model, $\forall~ J_{\gamma} >$ 0, is obtained. For example, in the limit, $J_r \rightarrow$ 0, the system reduces to two uncoupled antiferromagnetic spin-1/2 XXZ chains and can be analytically solved by using the Bethe ansatz \cite{bethe}. In this case, the ground state phase diagram is divided into a generic ferromagnetic phase for $\Delta < -1$, an \emph{XY}-phase for $-1<\Delta<1$, and a N\'eel ordered antiferromagnetic phase for $\Delta>1$. For strong ferromagnetic rung coupling, i.e., $J_r \rightarrow -\infty$, the ladder reduces to a spin-1 XXZ chain with weak interactions. In this limit, a predominant phase, apart from the ferromagnetic, \emph{XY}, and N\'eel phases, is the distinctly gapped Haldane phase \cite{spin1}. At the $\Delta/J_r \rightarrow \pm \infty$ limit, the ladder reduces to a pair of uncoupled Ising chains that give rise to a ferromagnetic or N{\'e}el phase. Depending on whether the coupling, $J_r\Delta$, is ferromagnetic or antiferromagnetic, these phases could be regular or \textit{stripe} ordered. This is discussed in more detail in Sec.~\ref{Phase}.

The main motivation of our work is to analyze the ground state phase diagram of the \textit{XXZ} ladder, given in Eq.~(\ref{hamiltonian}), by studying bipartite and multipartite entanglement transitions in the phase space. We concentrate on the \textit{XXZ} ladder with antiferromagnetic legs, and fix $J_{\gamma}$ = $J_l$, $\forall~ \gamma$, where $J_l$ is a positive quantity, and vary the inter-leg or rung coupling, $J_r$.
We begin by obtaining the ground states of this Hamiltonian in the $S^z$ = $0$ subspace, 
using the exact diagonalization method \cite{titpack}. We note that the characterization of genuine multipartite entanglement in large quantum spin systems requires explicit computation of quantum correlation across all possible bipartitions of the exact ground state. Such characterizations are computationally restricted in systems with large Hilbert spaces (cf.~\cite{RVB1,RVB2}), primarily due to the fact that the ground state cannot be computed for large system size, $N$.
This limits our investigation to moderately large spin ladders.
We investigate the variation of ground state properties with respect to the anisotropy parameter $\Delta$ and the rung-leg coupling ratio, $\alpha$ ($={J_{r}}/{J_{l}}$).

\begin{figure}[b]

\hspace{-1.5cm}\epsfig{figure =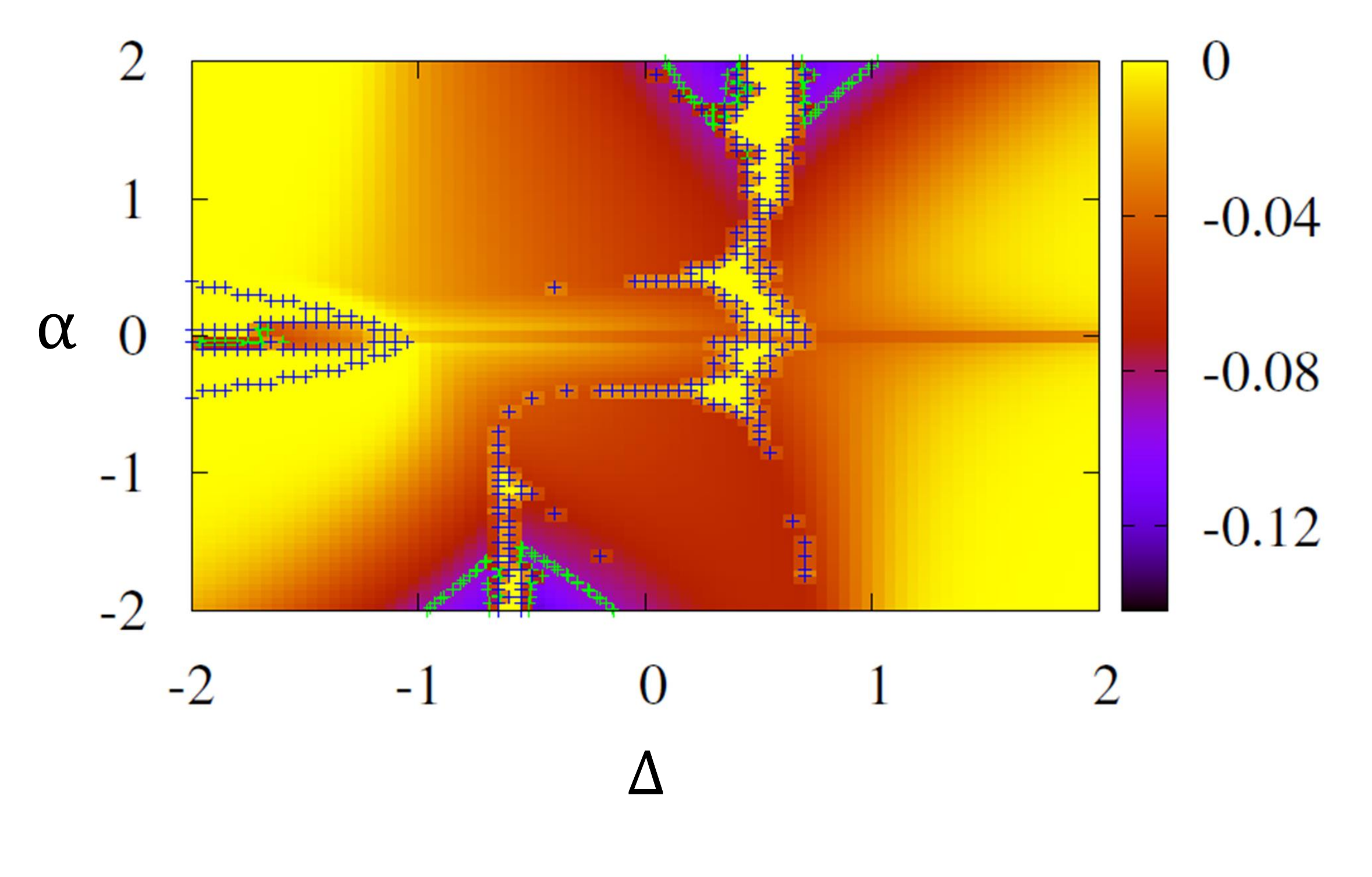, width=0.5\textwidth,angle=-0}\hspace{-1.5cm}
\caption{(Color online) Energy and the phase diagram of the spin-1/2 \emph{XXZ} ladder.
The plot shows variation of the excitation energy difference per spin, $\delta E/J_l$ = ($E_0 - E_1$)/$(J_lN)$, for a moderate-size system, in the $\alpha-\Delta$ plane, where $\alpha$ = ${J_{r}}/{J_{l}}$ is the rung-leg coupling ratio, and  $\Delta$ is the anisotropy constant. We set $J_\gamma$ = $J_{l}$ $\forall~\gamma$, and the number of spins is $N$ = $16$. Periodic boundary condition have been considered. The couplings along the legs are antiferromagnetic. All the quantities plotted are dimensionless.} 
\label{energy}
\end{figure}

We observe that certain conventional order parameters are able to exhibit specific parts of the phase diagram and identify specific phase boundaries in the $\alpha-\Delta$ plane \cite{strong, watanbe, hijii,zanardi}. For example, the energy gap, $\delta E$, which is the difference in energies between the ground and first excited states in moderately large systems, is able to anticipate
some of the phases of the model, as shown in Fig.~\ref{energy}.
%

In this context, the central question we address in this work is whether a distinctive pattern of entanglement can provide a more accurate insight about the phase boundaries of the system, even for moderately large spin systems.
%
To this end, we employ both NN bipartite entanglement, as quantified by concurrence ($\mathcal{Q}$) \cite{wooters}, and genuine multipartite entanglement measure, generalized geometric measure ($\mathcal{G}$) \cite{ggm} (cf. \cite{multi-ent,ggm2}), and analyze the transition of these quantities
in $\alpha-\Delta$ plane. Apart from entanglement,
%
the behavior of NN spin correlation functions, $\mathcal{C}^{\beta\beta'}(\rho_{ij})$ = $\langle S^{\beta}_i S^{\beta'}_{j}\rangle-\langle S^{\beta}_i \rangle \langle S^{\beta'}_{j}\rangle$, where $\beta,\beta' \in \{x,y,z\}$, along with other ground state properties such as magnetization, also provide important information in determining the different phase boundaries of the \emph{XXZ} ladder. Moreover, one should also note that, when the inter-rung interaction is antiferromagnetic, and there are an odd number of rungs, the end spins along the legs become frustrated due to the periodic boundary conditions. This plays an important role in the choice of minimum energy configuration and hence has a significant impact on the phase diagram. We will discuss this in detail in Sec.~\ref{Phase}

\section{Characterization of correlations in the $\alpha-\Delta$ plane}
{\label{properties}}
In this section, using the exact diagonalization method \cite{titpack}, we numerically calculate the variation of bipartite and multipartite entanglement measures, along with classical quantities, such as spin correlation functions, in the $\alpha-\Delta$ Cartesian plane, to identify and investigate the phase boundaries of the ground states of the spin-1/2 \emph{XXZ} ladder Hamiltonian, as defined in Eq.~\ref{hamiltonian}. As the Hamiltonian of interest  has rotational symmetry along the $z$-axis ($S^z$ invariant), all the computation have been carried out in $S^z=0$ subspace
(cf.~\cite{zanardi}).
%

\vspace{-0.4cm}
\subsection{Phase patterns from bipartite entanglement}

Let us first define concurrence, $\mathcal{Q}(\rho_{ij})$ \cite{wooters}, which we use as a measure to quantify bipartite entanglement in the model.
Given a two-qubit density matrix, $\rho_{ij}$, for any two sites of the ladder, the concurrence  of $\rho_{ij}$ is defined  as
\begin{equation}
\mathcal{Q}(\rho_{ij})=\max \{0, \lambda_1 - \lambda_2 - \lambda_3 - \lambda_4\},
\end{equation}
where $\lambda_i$'s are the square roots of the eigenvalues of the non-Hermitian  matrix, $\rho_{ij}{\tilde{\rho}_{ij}}$, arranged in decreasing order. Here
${\tilde{\rho}_{ij}}$ = $(\sigma_y \otimes \sigma_y) \rho^\ast_{ij} (\sigma_y \otimes \sigma_y)$ is the spin-flipped state of $\rho_{ij}$, with  $\rho^*_{ij}$ being the complex conjugate of the state in the computational basis. The physical interpretation of this definition stems from the connection of concurrence with the entanglement of formation, which quantifies the number of singlets required to create a bipartite quantum state by using only local quantum operations and classical communication \cite{wooters,wooters2}

Depending on the positions of the sites $i$ and $j$ in the ladder system, there are two relevant cases of NN bipartite entanglement --  1) when both the NN sites belong to the same leg, we get $\mathcal{Q}_{l}(\rho_{ij})$, which captures the bipartite entanglement between sites along the legs of the ladder; 2) each of the two NN sites belong to different legs, and the corresponding bipartite entanglement is denoted by $\mathcal{Q}_{r}(\rho _{ij})$, providing the entanglement behavior in the rungs of the ladder.

In Fig.~\ref{con_leg}, variation of $\mathcal{Q}_{l}(\rho_{ij})$ in the $\alpha-\Delta$ plane is exhibited for $N$ = 16 spins. We observe that for $\Delta <$ 0, the value of $\mathcal{Q}_{l}(\rho_{ij})$ remains low ($\mathcal{Q}_{l} <$ 0.2) or negligible ($\mathcal{Q}_{l} \approx$ 0) in significant regions of the parameter space. Hence, one can infer that the nonclassical correlations along the legs are relatively weak in these parameter regimes. On the other hand, in the region $\Delta >$ 0, the distribution of $\mathcal{Q}_{l}(\rho_{ij})$ efficiently detects the phase boundaries in the model and provides vital information about the nature of the ground state. For example, the variation of $\mathcal{Q}_{l}(\rho_{ij})$ in Fig.~\ref{con_leg} exhibits lines of transitions (entanglement peaks) at $\Delta$ = 1, $\alpha < 0$, and along $\alpha$ = 0, for $\Delta > 0$.  The same can also be observed at $\Delta \approx$ 1, $\alpha > 0$. These lines correspond to possible phase boundaries for the spin-1/2 \emph{XXZ} ladder \cite{hijii, zanardi}.
\begin{figure}[h]
\epsfig{figure = 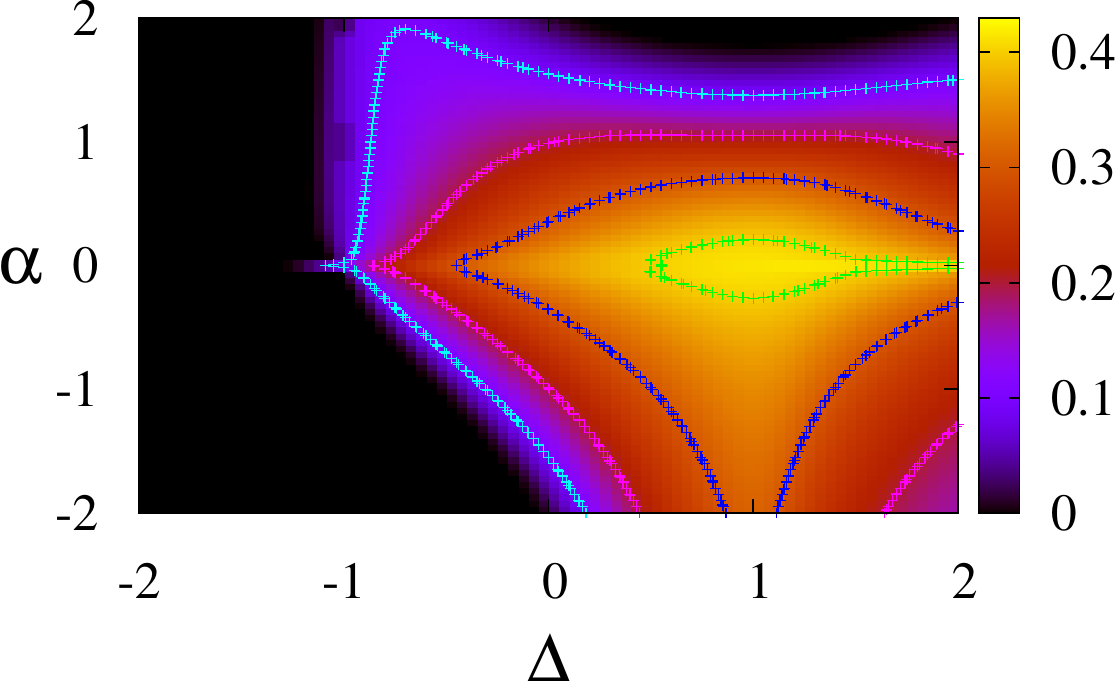, width=0.4\textwidth,angle=-0}
\caption{(Color online) 
Entanglement along NN leg sites of the spin-1/2 \emph{XXZ} ladder. The plot shows behavior of $\mathcal{Q}_{l}(\rho_{ij})$, in the $\alpha-\Delta$ plane. We choose $N=16$, so that we have an even numbers of rungs.
All other considerations for the model are the same as in Fig.~\ref{energy}. All quantities plotted are dimensionless, except the entanglement, which is in ebits.}
\label{con_leg}
\end{figure}

\begin{figure}[h]
\epsfig{figure = 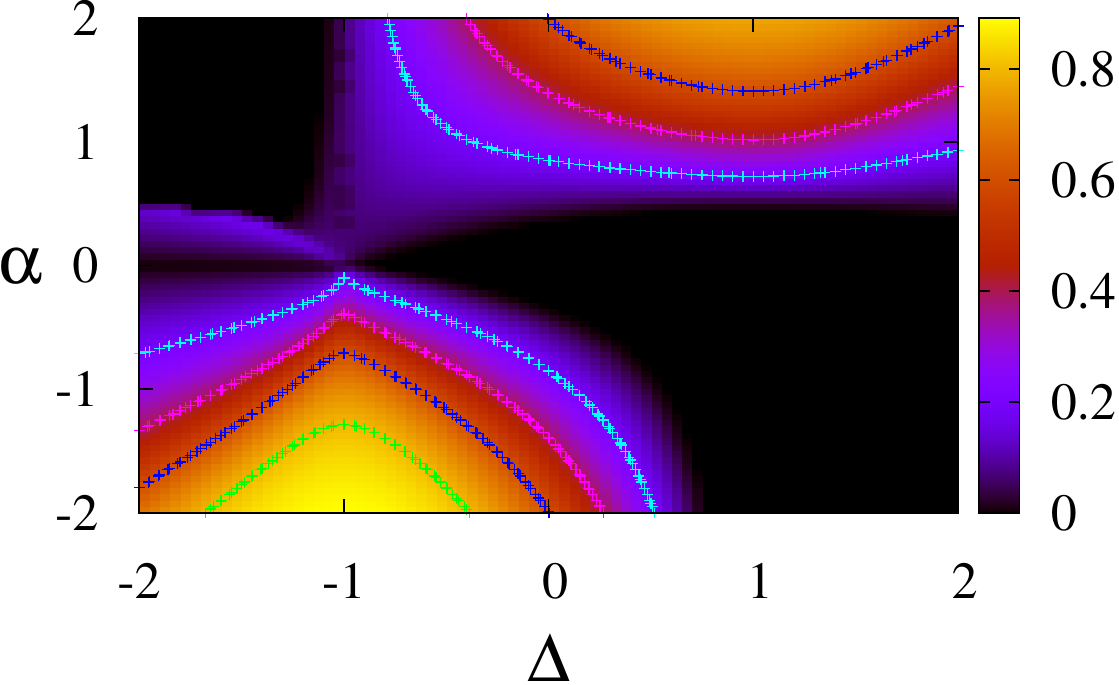, width=0.4\textwidth,angle=-0}
\caption{(Color online)
Entanglement along the rungs, $\mathcal{Q}_{r}(\rho_{ij})$, of the spin-1/2 \emph{XXZ} ladder
in the $\alpha-\Delta$ plane. All other considerations  are the same as in Fig.~\ref{con_leg}. All quantities plotted are dimensionless, except the entanglement which is in ebits.}
\label{con_rung}
\end{figure}

The concurrence along the rungs, $\mathcal{Q}_{r}(\rho_{ij})$, which is computed for the two-sites obtained by tracing out all the ladder sites except those belonging to any one of the rungs, exhibits quite a rich phase diagram as depicted in Fig.~\ref{con_rung}. This is intuitively expected as the spin-1/2 \emph{XXZ} ladder tends to undergo transitions between the typical phases of a spin-1 \emph{XXZ} chain, which are the gapped Haldane and rung-singlet phases, and the gapless XY phases. All these transitions are closely related to the entanglement behavior along the rungs of the spin-1/2 \emph{XXZ} ladder model.
A comparison with the plots obtained for
$\mathcal{Q}_{l}(\rho_{ij})$ (Fig.~\ref{con_leg})
leads us to realize that the behavior of $\mathcal{Q}_r(\rho_{ij})$, in the $\alpha-\Delta$ plane,
is complementary to the behavior of $\mathcal{Q}_l(\rho_{ij})$.
Specifically, there are regions in the $\alpha-\Delta$ plane where $\mathcal{Q}_r(\rho_{ij})$ possesses significantly value while $\mathcal{Q}_l(\rho_{ij})$ is negligible, and vice-versa.
Moreover, there is additional information about the phases that seem to emerge. For example, Fig.~\ref{con_rung} shows a distinct line of transition (entanglement peaks)  in the variation of $\mathcal{Q}_{r}(\rho_{ij})$ at $\Delta$ = $-$1, $\alpha < 0$. Additionally, similar to $\mathcal{Q}_{l}(\rho_{ij})$ in Fig.~\ref{con_leg}, the boundary at $\Delta \approx$ 1, $\alpha > 0$ is also exhibited, although the character of $\mathcal{Q}_r(\rho_{ij})$ and $\mathcal{Q}_l(\rho_{ij})$ is complementary. Figure~\ref{con_rung} shows an entanglement peak, as compared to an entanglement trough or valley in Fig.~\ref{con_leg}.
These transitions are expected phase boundaries of the spin-1/2 \emph{XXZ} ladder, and are discussed in greater detail in the following sections. If one compares Figs.~\ref{con_leg} and \ref{con_rung}, with the behavior of the excitation energy difference in Fig.~\ref{energy}, one observes that bipartite entanglement already provides a much diversified phase picture as compared to the traditional order parameters. 

Importantly, similar to Fig.~\ref{con_leg}, there exists a significant region in the $\alpha-\Delta$ parameter space, where the value of $\mathcal{Q}_{r}(\rho_{ij})$ remains vanishingly small. Moreover, in the range  $1/2\lesssim \alpha \lesssim2$ and $-2 \lesssim\Delta\lesssim -1$, both $\mathcal{Q}_{l}(\rho_{ij})$ and $\mathcal{Q}_{r}(\rho_{ij})$ remain negligible, implying absence of significant bipartite entanglement. We will show that the phase boundaries in these regions can be characterized using a genuine multiparty entanglement measure that accounts for the absence of local quantum correlations in the system.

\vspace{-0.4cm}
\subsection{Trends of genuine multipartite entanglement}

An $N$-party pure state is genuinely multiparty entangled if it is not separable across any bipartition.
The genuine multisite entanglement, of a quantum state $|\xi_N\rangle$, in possession of the parties $A_1, A_2, A_3,...A_N$, can be computed using the generalized geometric measure (GGM) \cite{ggm}. The GGM of an $N$-party quantum state is the optimized fidelity distance of the state from the set of all states that are not genuinely multiparty entangled.
More precisely, the GGM ($\mathcal{G}(|\xi_N\rangle)$) is defined as
\begin{equation}
\mathcal{G}(|\xi_N\rangle=1-\Lambda_{max}^2(|\xi_N\rangle),
\end{equation}   
where $\Lambda_{\max} (|\xi_N\rangle ) = \max | \langle \chi|\xi_N\rangle |$,
with the maximization being over all $N$-party quantum states 
$|\chi\rangle$
that are not genuinely multisite entangled.
\begin{figure*}[]
\subfigure[]{\includegraphics[width=7.2cm]{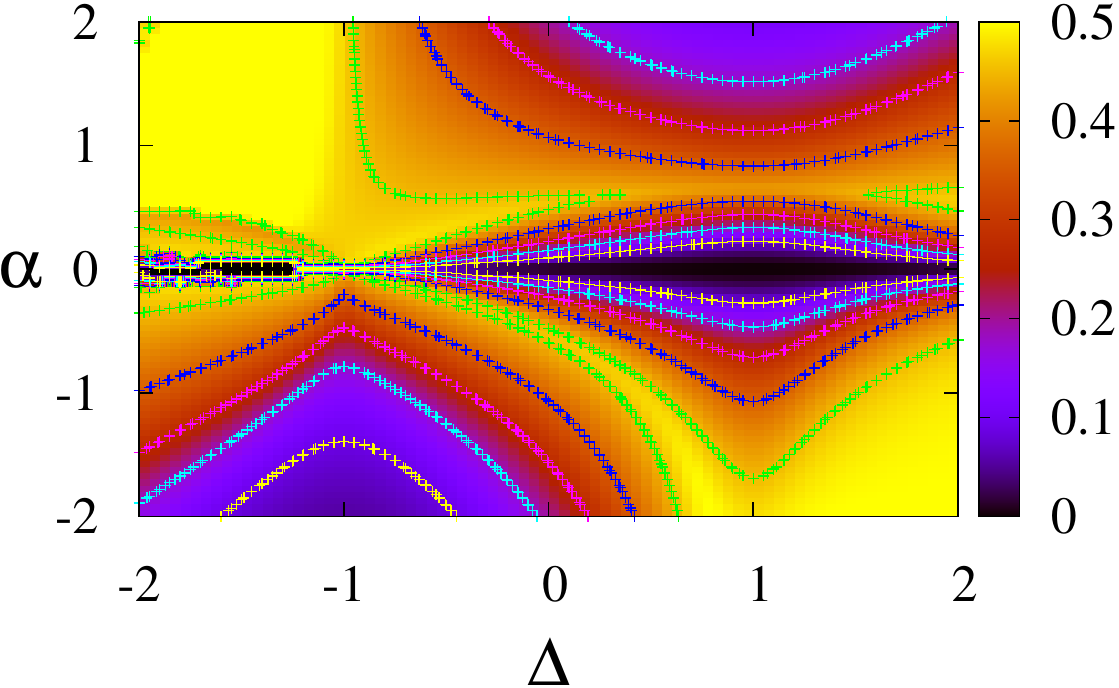}}\hspace{1cm}
\subfigure[]{\includegraphics[width=7.2cm]{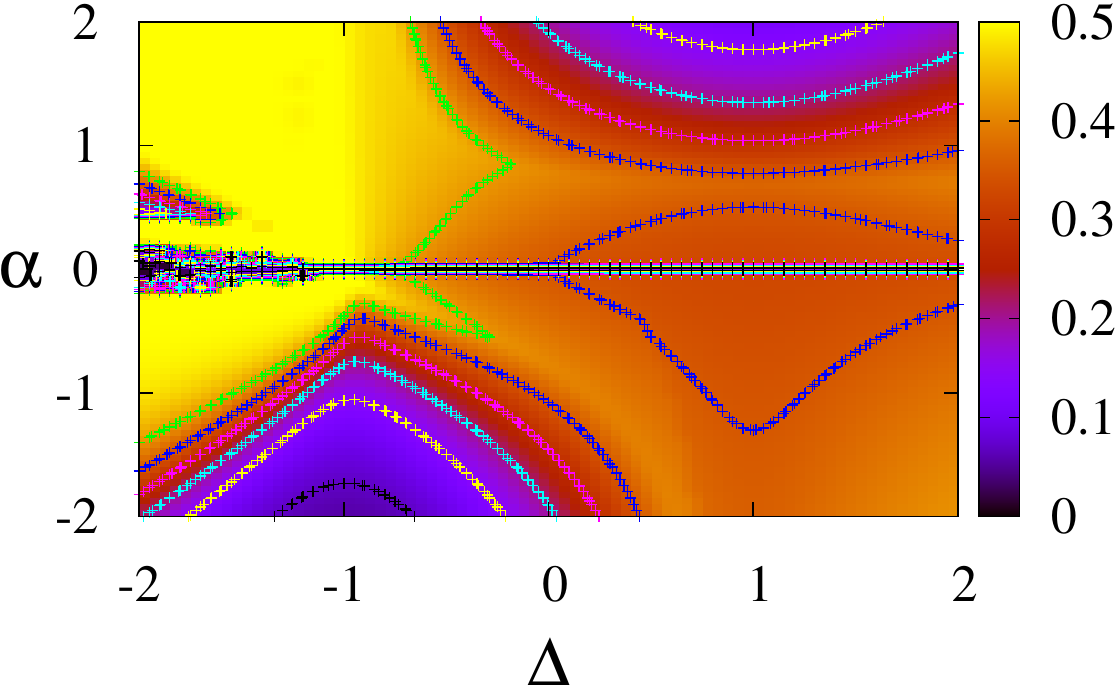}}
\caption{(Color online) Genuine multipartite entanglement of the spin-1/2 \emph{XXZ} ladder. The plots show variation of $\mathcal{G}$, in the $\alpha-\Delta$ plane for (a) $N=16$, so that there are an even number of rungs, and for (b) $N=18$, when there are an odd number of rungs. All other considerations are the same as in Fig.~\ref{energy}. The variation in the values of $\mathcal{G}$ between odd and even rung ladders, in the low rung-coupling regime, i.e., $J_r \approx 0$, is due to the spin frustration in the individual legs present in odd rung ladders with periodic boundary condition. However, the phase boundaries are identical. All quantities plotted are dimensionless.}
\label{fig_ggm}
\end{figure*}
Importantly, for pure quantum states, the GGM can be effectively computed using the relation
\begin{equation}
\mathcal{G} (|\xi_N \rangle ) =  1 - \max \{\lambda^2_{ A : B} |  A \cup  B = \{A_i\}_{i=1}^N,  A \cap  B = \emptyset\},
\label{GGM}
\end{equation}
where \(\lambda_{A:B}\) is  the maximal Schmidt coefficient of $|\xi_N\rangle$ in the bipartite split \(A: B\).

Interestingly, careful observation leads to the fact that the behavior of $\mathcal{G}$ is crucial in capturing the phase variation that is individually not attested by the bipartite entanglements, $\mathcal{Q}_{l}(\rho_{ij})$ and $\mathcal{Q}_{r}(\rho_{ij})$.
See Figs.~\ref{fig_ggm}(a) for $N$ = 16 and \ref{fig_ggm}(b) for $N$ = 18.
For example, the variation of $\mathcal{G}$ is complementary to the behavior we observe for bipartite entanglement along the rungs, $\mathcal{Q}_{r}(\rho_{ij})$, in the regions: i) $\alpha>0$, $\forall~\Delta$, and ii) $\alpha<0$, $\Delta <1$ (comparing with Fig.~\ref{con_rung}). However, in the region $\alpha<0$, $\Delta >1$, $\mathcal{G}$ highlights the phase variation not captured by $\mathcal{Q}_{r}(\rho_{ij})$, although shown by $\mathcal{Q}_{l}(\rho_{ij})$ (see Fig.~\ref{con_leg}). Moreover, there exists a wide region in parameter space for which the ground state of the system remains maximally genuine multiparty entangled, containing global entanglement in absence of short-range quantum correlations. Essentially, it can be seen that the region on the $\alpha-\Delta$ plane having negligible bipartite entanglement ($1 \lesssim \alpha \lesssim 2$ and  $-2 \lesssim\Delta\lesssim -1$), exhibits significant multipartite entanglement.
We note that such a study of the entanglement in this non-integrable model yields two important points -- 1) First, for moderately large ladders, the conventional critical indicators (e.g., see Fig.~\ref{energy}), such as energy, magnetization, and classical correlation functions, fail to capture the microscopic details of the many-body system and  essentially, cannot identify the sharp phase boundaries emanating from such changes. Therefore, one may argue, that for parameter regimes where the characterization of microscopic details are important, entanglement in general, and multipartite entanglement in particular, is a superior quantity to detect transitions than the conventional order parameters. 2) Secondly, there are several quantum information processing tasks in which bipartite as well as multipartite entanglement are essential ingredients. These results show that ground state phases of the spin-1/2 $XXZ$ ladder are potential substrates for realizable implementation of such tasks.

To supplement the investigation of different phases using entanglement, we also study the behavior of the spin correlation functions, particularly at the large values of $\Delta$ where Ising terms dominate. This allows us to obtain a more holistic characterization of the different phases of the ground state of the spin-1/2 \emph{XXZ} ladder.

\vspace{-0.4cm}
\subsection{Spin correlation function}
The pairwise classical correlation between any two sites in a many-body system, as quantified by the spin correlation function,
\begin{eqnarray}\label{cor_func}
\mathcal{C}^{\beta \gamma}(\rho_{ij})=\langle S^{\beta}_i S^{\gamma}_{j}\rangle-\langle S^{\beta}_i \rangle \langle S^{\gamma}_{j}\rangle,
\end{eqnarray}
where $\{\beta,\gamma\} \in\{x,y,z\}$, is an important order parameter that indicates quantum critical phenomena
\cite{sachdev, golden,sondhi}.
On the other hand, using the Lieb-Robinson bound, one can essentially show that for gapped systems,  the spin correlation function decays exponentially with the site distance, whereas for a gapless system, a polynomial decay is observed \cite{lieb}. In the spin-1/2 \emph{XXZ} ladder, presence of $\mathbb{Z}_2$ symmetry, and other considerations, ensure that the only non-zero two-site spin correlation functions of the ground state are $\langle S^{x}_i S^{x}_{j}\rangle$, $\langle S^{y}_i S^{y}_{j}\rangle$, and $\langle S^{z}_i S^{z}_{j}\rangle$. In the absence of any external field, all the single-site terms in Eq.~(\ref{cor_func}), $\langle S^{\beta}_i \rangle$ for $\beta \in \{x,y,z\}$, vanish. Hence, the correlation function reduces to $\mathcal{C}^{\beta\beta}(\rho_{ij})$ = $\langle S^{\beta}_i S^{\beta}_{j}\rangle$. 
Also, due to the isotropy along the $x$ and $y$ axes, we have $\mathcal{C}^{xx}(\rho_{ij})$ = $\mathcal{C}^{yy}(\rho_{ij})$.

The variation of two-site NN $\mathcal{C}^{xx}$ and $\mathcal{C}^{zz}$, along the legs and the rungs, in the $\alpha-\Delta$ plane,
does not effectively capture the phase boundaries for the \emph{XXZ} spin-1/2 ladder model. However, they provide a good estimate of the nature of the classical correlations in the various entangled phases of the model, which is a critical supplement in identifying the phase boundaries using entanglement. For example, the values of the functions $\mathcal{C}^{xx}$ and $\mathcal{C}^{zz}$ at the limiting values of $\Delta \rightarrow \pm \infty$
allows us to understand the onset of the striped ferromagnetic phase ($\Delta \ll 0$),
and the regular and striped N{\'e}el phases ($\Delta \gg 0$). These phases can be identified using traditional order parameters and have more definitive values of spin correlation functions.


A complete inference about the phase boundaries obtained from the trends of bipartite and multipartite entanglement, along with the behavior of spin correlation functions, is carried out in the succeeding sections.

\section{Detection of Phase boundaries in the spin$-\frac{1}{2}$ \emph{XXZ} ladder}
\label{Phase}
In this section, we analyze the phase variation of the  different quantum quantities in the $\alpha-\Delta$ plane, obtained in Sec.~\ref{properties}, to estimate the various ground state phases of the spin-1/2 \emph{XXZ} ladder.
We aim at providing a schematic diagram of the various phase boundaries that can be argued on the basis of the transition of these quantities in the $\alpha-\Delta$ plane.

\begin{figure}[h]
\begin{center}
\epsfig{figure = 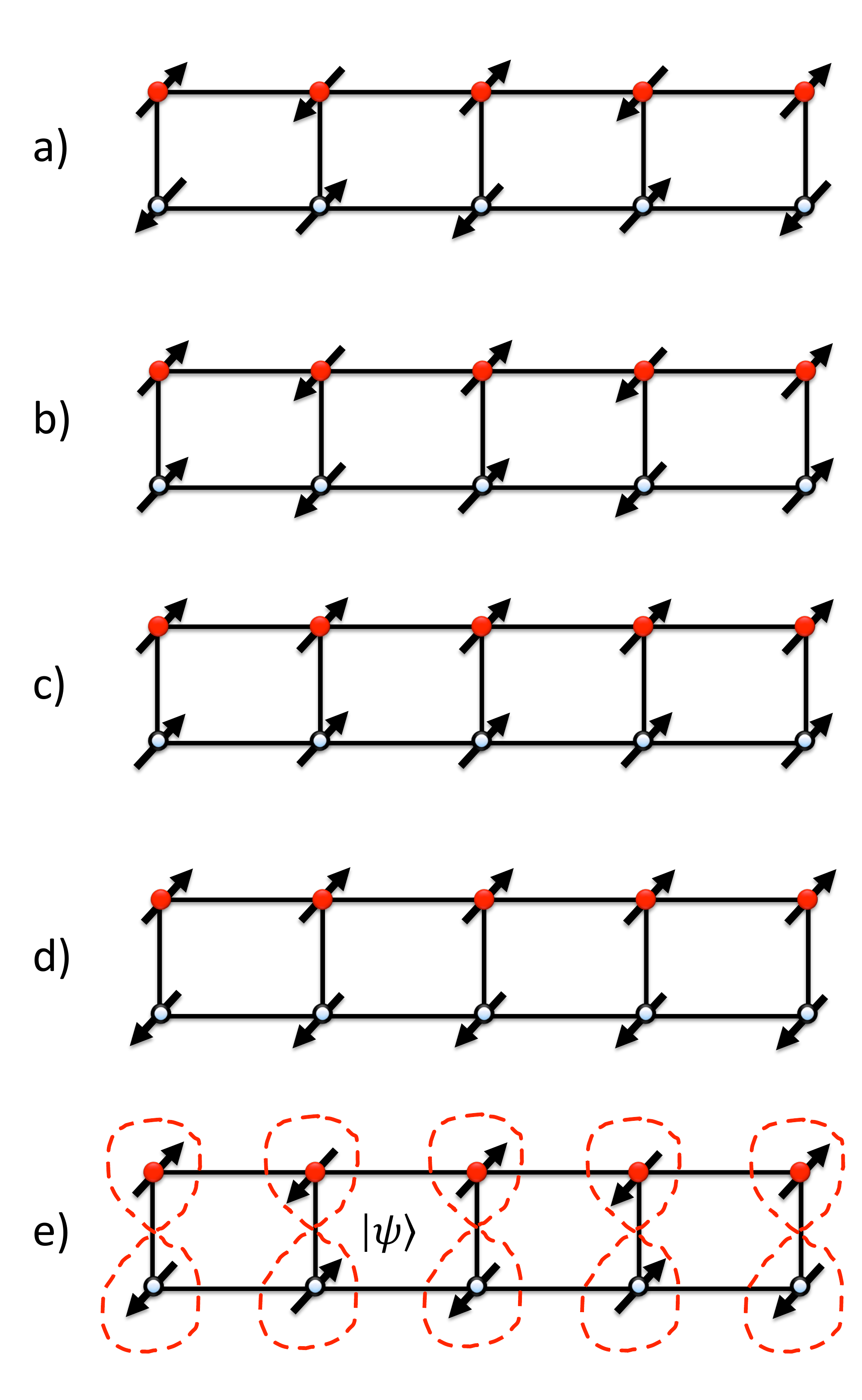, width=0.45\textwidth,angle=-0}
\caption{(Color online) Schematic diagrams of the ground state phases. The illustrations show the spin configurations in some of the potential ground states of the spin-1/2 \emph{XXZ} ladder:
(a) The regular ordered N\'eel state, (b) the stripe ordered N\'eel state, (c) the regular ordered ferromagnetic state, (d) the stripe ordered ferromagnetic state, and (e) the rung singlet state (the sites enclosed by the dotted lines represents a singlet, $|\psi\rangle$ = $\frac{1}{\sqrt{2}}(|\uparrow \downarrow\rangle-|\downarrow \uparrow\rangle)$). 
Here the tilted arrows represent up and down spins at the corresponding lattice sites. The blue and red circles correspond to the top and bottom legs of the ladder, respectively.}
\label{neel}
\end{center}
\end{figure}

We start with a description of the regular and stripe ordered ferromagnetic and antiferromagnetic ground states of the \emph{XXZ} ladder, as shown in Fig.~\ref{neel}. These ordered phases are predominantly exhibited in the limiting values of the anisotropy parameter, $\Delta$, i.e., $\Delta \rightarrow \pm \infty$, where the Ising-like interactions are dominant. In our study, for periodic ladders, we are able to highlight the onset of these ordered phases by analyzing the variation in the classical and quantum correlations. For large positive $\Delta$, the system tends to behave like an Ising ladder, with relatively large antiferromagnetic coupling ($J_\gamma \Delta$) between the leg spins.
Let us consider $\Delta \gtrsim$ 1.5.
Now, for $\alpha$ positive, the spins along the rungs are also coupled via an antiferromagnetic interaction, and a \textit{N\'eel} ordered ground state phase emerges (see Fig.~\ref{neel}(a)). However, for $\alpha$ negative, the spins along the rungs are  coupled via a ferromagnetic interaction, and we observe the \textit{striped N\'eel} ordered phase, shown in Fig.~\ref{neel}(b). This is supported by the behavior of the classical correlation functions and the entanglement.
In the regular ordered N\'eel phase, both $\mathcal{C}^{zz}_l$ and $\mathcal{C}^{zz}_r$, along the legs and rungs, respectively, are close to $-1$.
However, while low spin correlations are observed in the legs, along $x$ and $y$, they are close to $-1$ along the rungs (i.e., $\mathcal{C}^{xx}_l$ = $\mathcal{C}^{yy}_l \rightarrow$ 0, where $\mathcal{C}^{xx}_r$ = $\mathcal{C}^{yy}_r \rightarrow -1$).
This is supported by the fact that bipartite entanglement along the rungs increases, while it vanishes along the legs, as $\alpha$ increases in the ordered N\'eel phase.
For instance, $\mathcal{Q}_l \approx$ 0.4 near $\alpha=0$ and vanishes as $\alpha$ increases. On the other hand, $\mathcal{Q}_r$ is maximum and $\approx0.8$ for large $\alpha$. See Figs.~\ref{con_leg} and \ref{con_rung}. Hence, as $\alpha$ increases, the ladder is more entangled along the rungs, and tends towards the rung-singlet ground state.
In contrast, for the striped N\'eel phase, $\mathcal{C}^{zz}_l \approx -$1, but $\mathcal{C}^{zz}_r \approx$ 1 ($\mathcal{C}^{xx}_l \neq $0, $\mathcal{C}^{xx}_r \approx$0), implying ferromagnetic arrangement of NN spins along the rungs. In this region, we observe finite $\mathcal{Q}_l$, although $\mathcal{Q}_r$ = 0, as supported by the behavior spin correlation functions.
As observed in Fig.~\ref{fig_ggm}, the striped N\'eel phase possesses high genuine multipartite entanglement while low GGM is observed in the ordered N\'eel phase. These two phases are demarcated by a vanishing GGM value along the line $\alpha$ = 0.

\begin{figure}[t]
\epsfig{figure =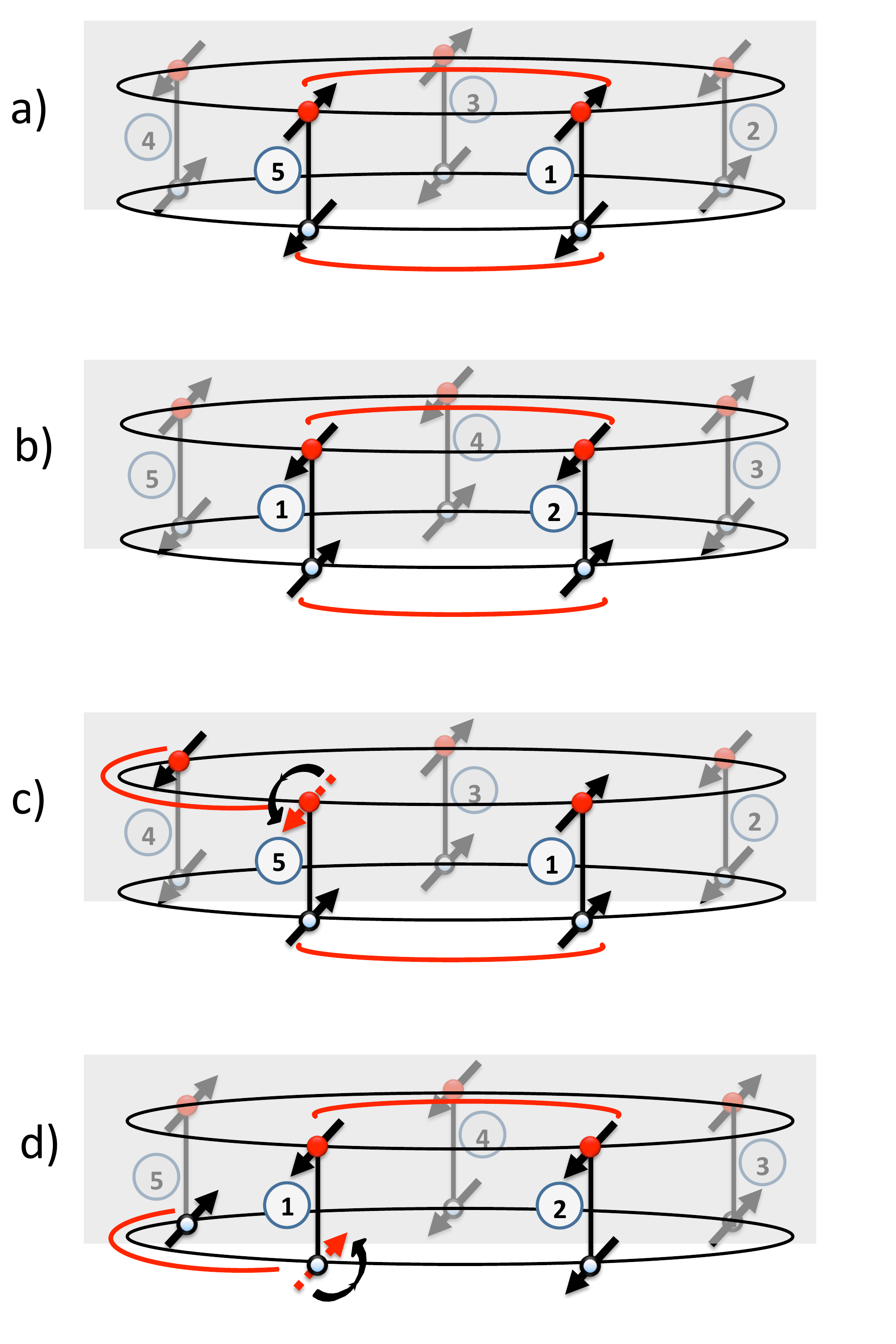, width=0.45\textwidth,angle=-0}
\caption{(Color online) Schematic diagrams showing frustration in the spin chains. In the illustrations, we show two examples of the many competing spin configurations leading to frustration between the end spins along the legs, viz. in (a)-(b) a regular ordered N{\'e}el state, and (c)-(d) a stripe ordered N{\'e}el state. The non-optimal spin alignment in the periodic lattice leading to frustration between the end spins of each of the legs is shown by curved red lines (see also Figs. \ref{neel}(a) and (b)). However,  to satisfy $S^z$ = 0, one of the end spins flips in the striped N{\'e}el state, which is shown by black curved arrows at site 5-(top) for (c) and 1-(bottom) for (d). The flipped spins at these sites are marked in red (depicted using broken lines). We note that the five lattice sites are labelled (1)-(5), with a periodic boundary, and the lattices in panels (b) and (d) are rotated by a rung to the right, along the vertical axis perpendicular to the radius, in comparison to the lattices in (a) and (c).
}
\label{frustu}
\end{figure}

Interestingly, \textit{$S^z$ invariance} and \textit{frustration} play important roles in the ground state properties of periodic ladders. For ladders with an \textit{even} number of spins on each leg (say, $N$ = 16), the regular and stripe ordered N{\'e}el phases are $S_z$ = 0 states, and have no frustration, i.e,
the N{\'e}el state containing an even number of spins is the minimum energy state.
However, for periodic ladders with an \textit{odd} number of spins on each leg (say, $N$ = 18), the N{\'e}el phase is frustrated.
For example, Fig.~\ref{frustu}(a) shows that for the periodic N{\'e}el ladder state with $N$ = 10, the first (site 1) and last (site 5) rung sites of the two legs are ``not antiferromagnetically'' aligned, leading to increase in energy in comparison to the fictitious situation where all interaction terms are simultaneously minimized. As shown in Fig.~\ref{frustu}(b), competing configurations may arise, as non-aligned spins may occur between other NN rungs (for instance, rung sites 1 and 2, in Fig.~\ref{frustu}(b)). Existence of competing ground state configurations leads to frustration in the system, which increases possible superpositions in the ground state and may result in relatively high and stable entanglement values.  
%
%
Moreover, for odd number of spins, the striped N{\'e}el phase is not only frustrated, but has $S_z \neq$ 0 (in fact $S_z$ = 1 in Fig.~\ref{neel}(b)). The competing configurations leading to frustration in this state are shown in Fig.~\ref{frustu}(c)-(d). Additionally, to satisfy the $S_z$ = 0 constraint, the striped N{\'e}el phase must undergo a spin flip at a single site (shown as site 5-(top) and site 1-(bottom) in Figs.\ref{frustu}(c) and (d), respectively). Hence, at least one NN rung spin-pair is antiferromagnetically paired (rest of the rungs have ferromagnetically aligned paired spins), leading to one NN spin-pair with ferromagnetic interaction on each leg, which relatively lowers the genuine multisite entanglement by breaking the N{\'e}el ordering in the legs.
This is observed by the lower values of $\mathcal{G}$ in these limits, compared to ladders with an even number of spins in each leg . See Figs.~\ref{fig_ggm}(a) and (b).
Hence, the effect of frustration in the N{\'e}el phase of ladders, with odd number of spins on each leg, is that the variation of entanglement is not as diverse as compared to ladders with even number of spins on each leg.

On the other hand, at $\alpha \approx$ 0 and $\Delta >$ 0, the frustration in the odd case leads to the ground state to having considerably higher and stable entanglement, as compared to the even one, which indeed has rich variation of $\mathcal{G}$ along the $\alpha$ = 0 line ($\Delta >$ 0) (see Fig.~\ref{fig_ggm}(a)). Here, by ``stable" entanglement, we mean that the value of the entanglement in the relevant parameter space region remains virtually unchanged.

We now consider the region with relatively large negative $\Delta$ ($< -1.5$), where the system tends to behave like an  Ising ladder with ferromagnetic interaction between the leg spins.
For $\alpha >$ 0, the interaction between the rung spins is ferromagnetic ($J_r\Delta <$ 0). This should ideally lead to the regular \textit{ferromagnetic} ordered phase, as shown in Fig.~\ref{neel}(c). However, this phase is not observed in our analysis as it has $S^z \neq$ 0. Instead, numerical studies show that the phase corresponds to variations of the \textit{striped ferromagnetic} states, shown in Fig.~\ref{neel}(d), which has $S^z$ = 0, with ferromagnetic coupling on each leg. A distinctive feature of the region is that all bipartite entanglement is zero, along both legs and rungs, but the phase has maximum genuine multipartite entanglement.
In Ref.~\cite{zanardi},
it has been suggested that for the range of parameters, $1 \lesssim \alpha \lesssim 2$ and $-2 \lesssim \Delta \lesssim -1$,  the ground state of the system is likely to be in a rung singlet phase \cite{zanardi}, based on analysis of selective entanglement entropy. However, our characterization of the microscopic properties corresponding to this phase do not support this fact. From the  variation of $\mathcal{Q}^{r}$ it can be clearly seen that within the mentioned range of parameters, the bipartite entanglement across the rungs turns out to be vanishingly small, which strongly contradicts the presence of rung singlet phase. However, from the analysis of $\mathcal{G}$, the phase is likely to have very high amount of genuine multiparty entanglement, making it closer to a possible superposition of striped ferromagnetic states.

For $\alpha <$ 0, the rung-spin coupling, $J_r \Delta$, is positive, and hence the interaction along the rungs is antiferromagnetic. The ground state phase is  \textit{striped ferromagnetic} as expected. This is characterized by $\mathcal{C}^{zz}_l \approx $1 and $\mathcal{C}^{zz}_r \approx -$1. The spin correlations $\mathcal{C}^{xx}$ ($\mathcal{C}^{yy}$) is negligible along the legs, but finite along rungs. The phase has states with finite entanglement along the rungs, which increases with decreasing $\alpha$, but is nonexistent along the legs. The phase is thus characterized by a region of decreasing genuine multipartite entanglement, which goes to zero as $\alpha$ decreases (see Fig.~\ref{fig_ggm}). However, we note that for odd rung ladders, and for $0.1<\alpha<-1$, the phase is more akin to the striped ferromagnetic phase observed for $\alpha>0$, characterized by maximum $\mathcal{G}$ (see Fig.~\ref{fig_ggm}(b)).

In the $\alpha-\Delta$ plane, the region defined by  0 $\lesssim \Delta <$ 2, and $\alpha\gtrsim$ 1.5 ($J_r > J_l$), is characterized by sharply increasing NN bipartite entanglement along the rungs, and negligible entanglement along the legs ($\mathcal{Q}_l \approx$) 0. Moreover, this region is also characterized by vanishingly low genuine multisite entanglement, $\mathcal{G} \approx$ 0.
As a consequence, the ground state properties of the system turns out to be close to the \emph{dimer} phase, as supported by the work in Ref.~\cite{himadri_new}. However as $\alpha$ increases, the ground state phase of the system in this region is close to the product of highly entangled pairs of spins along the rungs, giving rise to the \emph{rung singlet}, as shown in Fig.~\ref{neel}~(e). The rung singlet phase is also highlighted by a demarcated finite value region in the $\delta E$ phase diagram in Fig.~\ref{energy}, which corresponds to the valence bond solid phase. The presence of the rung singlet phase is clearly seen in the phase diagram of $\mathcal{G}$ in Fig.~\ref{fig_ggm}.
Another important phase that arises in the region, defined by $-1<\Delta<0$ and $\alpha \lesssim -1$ is the  \textit{XY} phase having high value of $\mathcal{Q}_r$, with negligible values of $\mathcal{Q}_l$ and $\mathcal{G}$. In this region, $|J_r| > J_l$ and $|\alpha| > \Delta$, leading to stronger \emph{XY} interactions. This is characterized by entanglement features similar to the rung singlet phase. The phase boundaries are well-demarcated by the variation of $\mathcal{G}$ in the $\alpha-\Delta$ plane, as shown in Fig.~\ref{fig_ggm}. For alternate expositions for the rung singlet and $XY$ phases, see \cite{hijii} and references therein. Other phases, highlighted by the genuine multipartite entanglement corresponds to the phase region between the $XY$ and striped N\'eel phases. Several studies have shown this region to mark the \textit{Haldane} phase.
In our analysis, this phase is marked by negligible entanglement along the rungs, with finite bipartite entanglement along the legs. The region is also marked with high $\mathcal{G}$ in the even  rung ladders (see Fig.~\ref{fig_ggm}(a)), with a distinct boundary separating it from the striped N\'eel phase, which also has high values of the generalized geometric measure.
%
An important point from our analysis is the implication that the phase boundary between the \emph{XY} and the Haldane phases extends to the $\Delta > 0$ region.  Though this has been observed in several studies~\cite{strong, f4,zanardi}, others have reported  $\Delta=0$  as the possible phase boundary \cite{narushima, legeza}.

\begin{figure}[t]
\epsfig{figure = 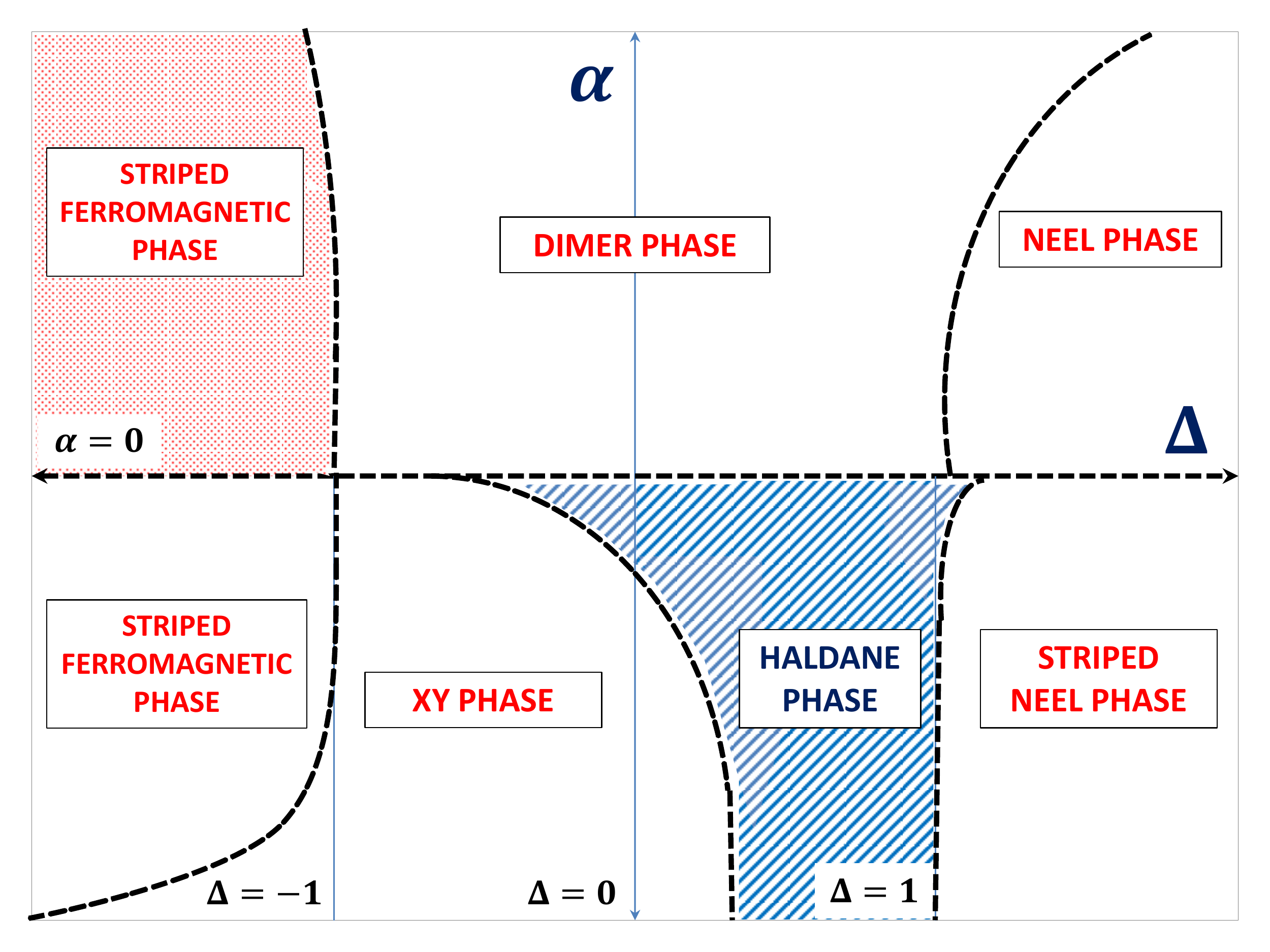, width=0.48\textwidth,angle=-0}
\caption{(Color online) Phase diagram of the spin-1/2 \emph{XXZ} ladder. An illustration is given here of the possible ground state phase diagram for the quantum spin-1/2 \emph{XXZ} ladder, in the $\alpha$-$\Delta$ plane, using the variation of bipartite and multipartite quantum entanglement. The phase diagram is consistent with those obtained from previous numerical studies \cite{strong, watanbe, hijii,zanardi}. The phase boundaries are indicated through black-dashed lines and the important phases, which are observed by analyzing the classical and quantum correlations, are labelled in red. The \emph{striped ferromagnetic} phase, which is observed in place of the expected ferromagnetic phase due to the $S^z$ invariance of the ground state, is shown by the red-dotted region in the $\alpha$-$\Delta$ plane. Moreover, we also indicate the expected \emph{Haldane} phase, which is marked by blue-dashed lines.}
\label{fig_phase}
\end{figure}

So from the above analysis, we are well-equipped to provide a schematic phase diagram for the ground state phases of the moderate-size spin-1/2 \emph{XXZ} ladder system, in the $\alpha-\Delta$ plane. The phase boundaries are between the various phases, namely a) the regular and striped N\'eel phases, b) the striped ferromagnetic phase, c) the rung-singlet phase, d) the \textit{XY}  phase, and e) the Haldane phase. The boundaries between these phases are primarily drawn by observing the transition lines of bipartite or multipartite entanglement or both. In most of the cases, the transition lines are exhibited by regions corresponding to entanglement peaks or troughs (valleys). For example, the transition between the striped ferromagnetic and $XY$ phases in the $\alpha <$ 0 and $\Delta <$ 0 region, is characterized by a $\mathcal{Q}_r$ peak along the $\alpha$ = $-1$ line. This is strengthened by the variation of $\mathcal{G}$, which shows clear troughs along the $\alpha$ = $-1$ phase boundary. Similar analyses of entanglement variation allows us to have a distinct picture of the different phases of the spin-1/2 \emph{XXZ} ladder.
We note that entanglement does not provide us a well-defined order to study the phase properties of the system, and we instead rely on the variation of entanglement to understand the phase boundaries. The different phases are named after the ordered states that dominate the corresponding parameter regime, as known from previous studies on the model. We do not imply that all states in a phase have some definite relation in terms of entanglement.

An illustrative phase diagram is given in Fig.~\ref{fig_phase}, showing the different phases described in the above analyses, with approximate phase boundaries.

\section{Spin$-\frac{1}{2}$ \emph{XXZ} ladder with ferromagnetic legs}
\label{ferro}

Until this point in our study, we have considered antiferromagnetic couplings along the legs of the ladder, i.e., $J_\gamma >$ 0. As mentioned earlier, coupled antiferromagnetic spin chains play a vital role in the study of materials exhibiting high-$T_c$ superconductivity and other strongly-correlated phenomena \cite{high_tc,karen}. However, spin-1/2 ladders with ferromagnetic interaction along the legs have also received considerable attention in the last few years \cite{f1,f2,f3,f4}, particularly in the design and study of synthetic ferromagnetic chains using Cu$^{2+}$ compounds \cite{f1}. Recent studies have also used theoretical approaches to study the phase structure of these systems \cite{f3,f4}.
Moreover, advances in material science and ultracold atoms may allow physicists to design and simulate ferromagnetic spin-1/2 ladders in future.

\begin{figure}[h]
\begin{center}
\epsfig{figure = 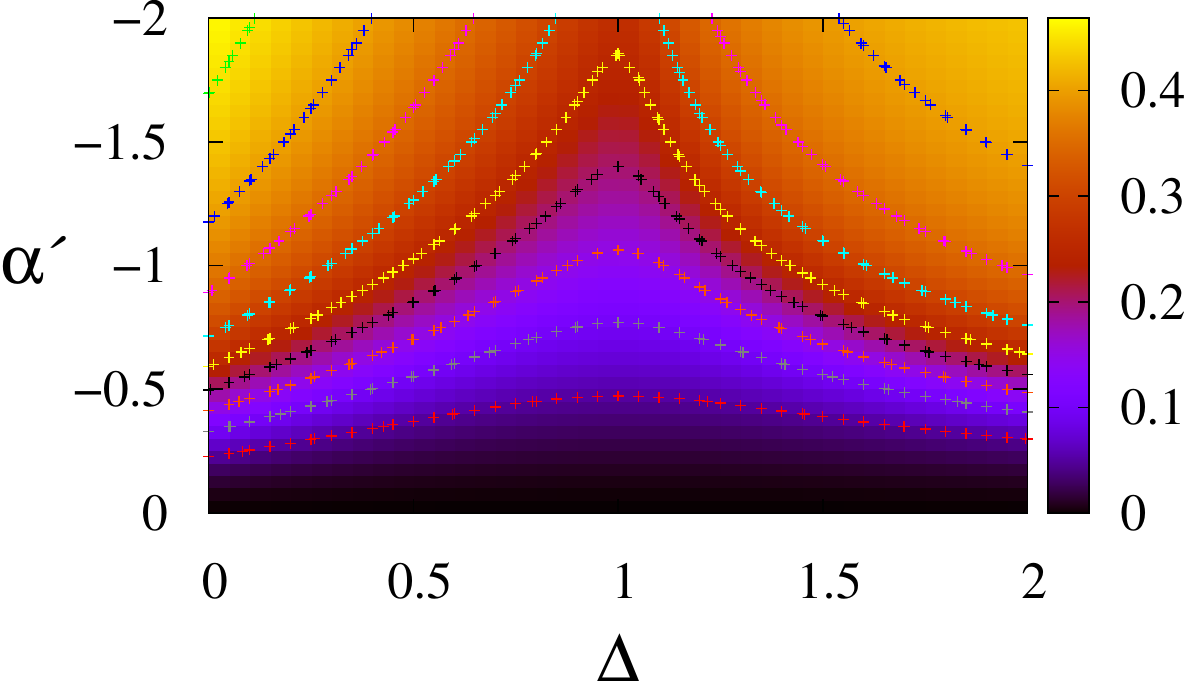, width=0.4\textwidth,angle=-0}
\caption{(Color online) Genuine multipartite entanglement in in spin-$1/2$ \emph{XXZ} ladders with ferromagnetic legs. Variation of $\mathcal{G}$, in the $\alpha^\prime-\Delta$ plane, where $\alpha^\prime$ = ${J_{l}}/{J_{r}}$ is the leg-rung coupling ratio, and  $\Delta$ is the anisotropy constant along the $z$ direction, for a spin-1/2 \emph{XXZ} ladder, with ferromagnetic coupling along the legs, such that $J_l <$ 0 and $\Delta >$ 0. The number of spins is $N$ = $14$. Periodic boundary condition is considered for the ladder. All quantities plotted are dimensionless.}
\label{fig_fggm}
\end{center}
\end{figure}

In the previous sections, we showed how the variation of bipartite and multipartite entanglement of the ground state, can reliably highlight the phase boundaries of a moderately large spin-$1/2$ \emph{XXZ} ladder, with antiferromagnetic legs. The approach can also be applied to spin ladders with ferromagnetic interaction along the legs, i.e., $J_\gamma <$ 0, $\forall~\gamma$, and $\Delta >$ 0. Investigation of the variation of genuine multipartite entanglement in the $\alpha^\prime-\Delta$ plane, where $\alpha^\prime$ = ${J_{l}}/{J_{r}}$, in Fig~\ref{fig_fggm}, shows that $\mathcal{G}$ is able to highlight the important phase boundaries as obtained in known studies based on effective field theories (cf.~Fig.~2 in \cite{f3}). For low $\alpha^\prime$, around $\Delta$ = 1 (isotropic case), a \textit{disordered singlet} ground state phase is observed, characterized by $\mathcal{G} \approx$ 0, which vanishes as $\alpha^\prime$ increases. To the right of the disordered singlet phase, for $\Delta >$ 1 and high $\alpha^\prime$, mean field studies reveal an \emph{ordered N{\'e}el} phase, and we find that this phase has relatively high $\mathcal{G}$. This is comparable to the ordered N{\'e}el phase observed in antiferromagnetic chains in Fig.~\ref{fig_ggm}. The region $\Delta <$ 1 and high $\alpha^\prime$, corresponds to the Kosterlitz-Thouless \emph{quasi-long-range order} (LRO) phase \cite{f3}. The behavior of $\mathcal{G}$ in this phase is similar to that in the N{\'e}el phase
and hence one cannot distinguish between the two phases using genuine multipartite entanglement. However, the variation of $\mathcal{G}$ in the singlet disordered phase clearly separates the LRO and N\'eel phase, which is consistent with known results.


\begin{figure}[h]
\epsfig{figure = 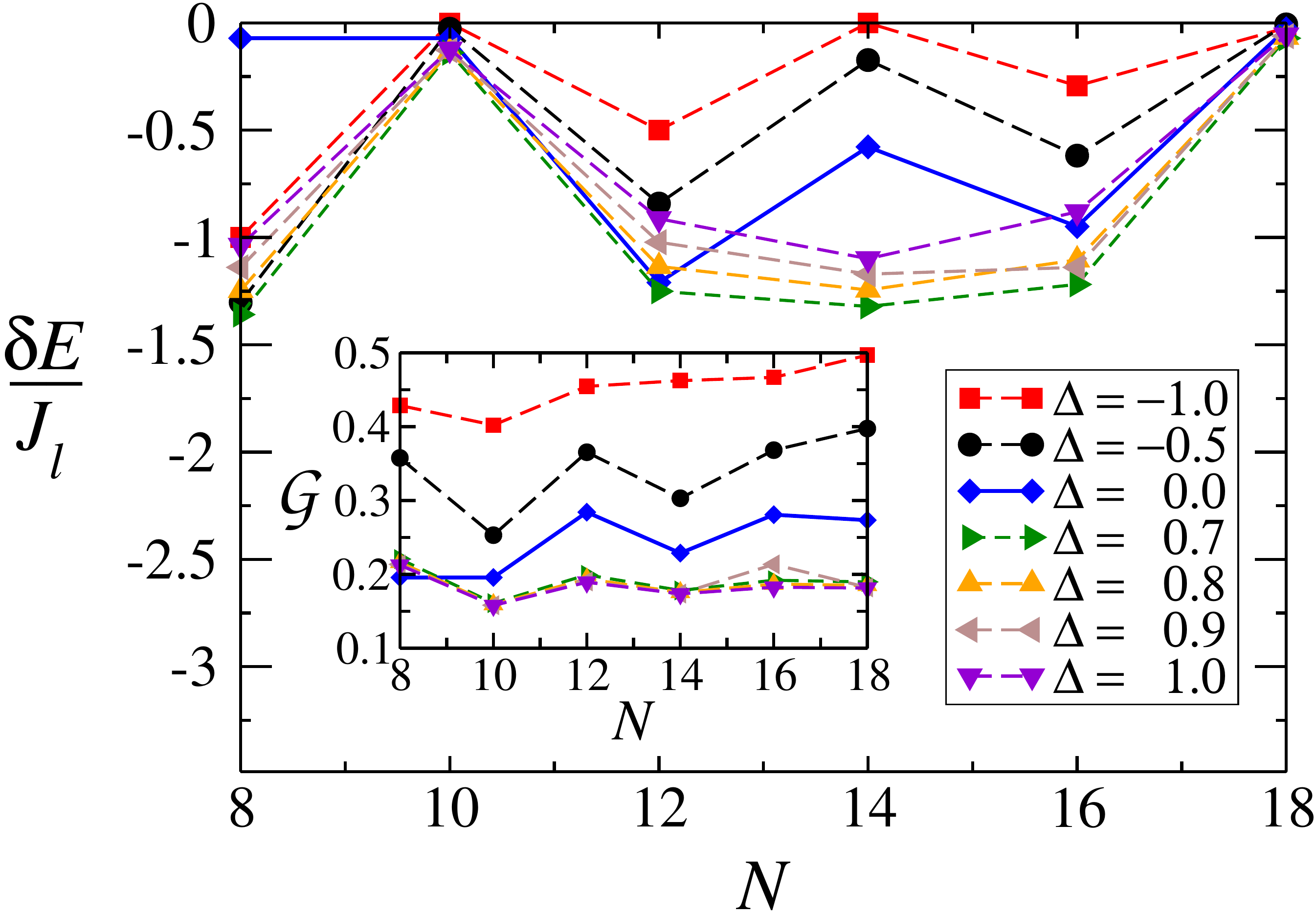, width=0.4\textwidth,angle=-0}
\caption{(Color online) Scaling of energy and genuine multiparty entanglement (inset) with increasing number of spins, in the spin-1/2 \emph{XXZ} ladder. The scaling of  $\delta E/J_l$ is almost complementary to $\mathcal{G}$. After $\Delta\approx$ 0.5, the lines in the ($\delta E/J_l,N$) plot for different values of $\Delta$, come close to each other, thus signaling identical scaling, independent of $\Delta$. The genuine multiparty entanglement as quantified by $\mathcal{G}$, also has identical scaling, after a certain value of the anisotropy parameter $(\Delta\approx 0.5)$. Here we consider $\alpha$ = ${J_{r}}/{J_{l}}$ = $1.5$. All the quantities plotted are dimensionless.}
\label{scaling}
\end{figure}

\section{Scaling of energy and multiparty entanglement}
\label{scale}

An important aspect of the study of the ground state phases of moderately large spin-1/2 \emph{XXZ} ladders is that the global entanglement properties of the state, such as genuine multiparty entanglement can be computed using exact diagonalization methods.
Now from our analysis, and the corresponding phase diagram, it is clear that conventional order parameters associated with critical phenomena, such as the excitation energy difference  ($\delta E/J_l$) in Fig.~\ref{energy}, can provide a coarse picture of the phase boundaries in moderately large systems. However, a more detailed phase diagram emerges by analyzing the quantifiers of microscopic quantum correlation properties of the system, such as the genuine multisite entanglement ($\mathcal{G}$).
An interesting question that arises is how the quantities $\delta E/J_l$ and $\mathcal{G}$ behave as the system-size of the ladder is increased.
In this regard, we perform a scaling analysis for both the quantities, to study their variations with increasing number of spins, $N$. This allows us to understand how these quantities extend to the thermodynamic limit, and whether the correspondence, which allows them to be used as suitable figures of merit in studying the phase properties of the ground state of a spin-1/2 \emph{XXZ} ladder, is broken.\

Figure~\ref{scaling} essentially depicts the way both the quantities scale with increasing size of the ladder. The notable point is that $\delta E/J_l$ and $\mathcal{G}$ exhibit almost complementary scaling with the system size, $N$. After a specific value of the anisotropy constant $\Delta$, the scalings are independent of $\Delta$, i.e., all the lines corresponding to different $\Delta$ values tend to overlap, signaling $\Delta$-independent scaling.
Interestingly, this shows that in the asymptotic limit, in parameter regimes where the system has a finite difference in excitation energy, both $\delta E/J_l$ and $\mathcal{G}$ can be useful physical quantities to detect potential phase transitions. 

\section{Conclusion}
\label{conclusion}
To summarize, in our work, we have considered moderately large quantum spin-1/2 \emph{XXZ} ladders, with both antiferromagnetic and ferromagnetic legs, and investigated the various ground state phases using
the quantum correlations measures, concurrence and generalized geometric measure, which capture the bipartite and the genuine multipartite entanglement in the system, respectively. We observe that the variation of entanglement, in the $\alpha-\Delta$ parameter space, provides a more diverse and distinct phase diagram, which is consistent with known results, as compared to conventional physical quantities such as the spin correlation functions.
%
Using the trends of bipartite as well as multipartite entanglement, computed for moderately sized systems of the nonintegrable model, we are able to highlight the possible phase diagram, which is in good agreement with known results obtained from different methods.
Additionally, we performed the scaling analysis, which shows that the macroscopic observable $\delta E$ and the quantifier of multiparty quantum correlations, $\mathcal{G}$, exhibit a close correspondence that seems to persist even in the asymptotic limit.
Our results provide an insight into the existence of different phases in the spin-1/2 \emph{XXZ} ladders, and enhance our understanding of quantum and classical correlation properties of their ground state phases. Moreover, high entanglement content in the system, of both bipartite and multipartite varieties, allow the possibility of its use in the implementation of quantum information protocols.

\end{document}